\input harvmac
\input epsf
\noblackbox

\newcount\figno

\figno=0
\def\fig#1#2#3{
\par\begingroup\parindent=0pt\leftskip=1cm\rightskip=1cm\parindent=0pt
\baselineskip=11pt \global\advance\figno by 1 \midinsert
\epsfxsize=#3 \centerline{\epsfbox{#2}} \vskip 12pt
\centerline{{\bf Figure \the\figno :}{\it ~~ #1}}\par
\endinsert\endgroup\par}
\def\figlabel#1{\xdef#1{\the\figno}}
\def\pano{\par\noindent}

\font\cmss=cmss10
\font\cmsss=cmss10 at 7pt

\def\rlx{\relax\leavevmode}
\def\inbar{\vrule height1.5ex width.4pt depth0pt}
\def\IC{\relax\,\hbox{$\inbar\kern-.3em{\rm C}$}}
\def\IR{\relax{\rm I\kern-.18em R}}
\def\IN{\relax{\rm I\kern-.18em N}}
\def\IP{\relax{\rm I\kern-.18em P}}
\def\frac#1#2{{#1 \over #2}}
\def\ZZ{\rlx\leavevmode\ifmmode\mathchoice{\hbox{\cmss Z\kern-.4em Z}}
 {\hbox{\cmss Z\kern-.4em Z}}{\lower.9pt\hbox{\cmsss Z\kern-.36em Z}}
 {\lower1.2pt\hbox{\cmsss Z\kern-.36em Z}}\else{\cmss Z\kern-.4em Z}\fi}

\def\narrowplus{\kern -.04truein + \kern -.03truein}
\def\narrowminus{- \kern -.04truein}
\def\narrowminussub{\kern -.02truein - \kern -.01truein}

\def\a{\alpha}

\def\o#1{\overline{#1}}

\def\ra{\rangle}



\lref\SchellekensAM{
A.~N.~Schellekens and S.~Yankielowicz,
``Extended Chiral Algebras And Modular Invariant Partition Functions,''
Nucl.\ Phys.\ B {\bf 327}, 673 (1989)\semi
A.~N.~Schellekens and S.~Yankielowicz,
``Modular Invariants From Simple Currents: An Explicit Proof,''
Phys.\ Lett.\ B {\bf 227}, 387 (1989)\semi
A.~N.~Schellekens and S.~Yankielowicz,
``New Modular Invariants For N=2 Tensor Products And Four-Dimensional
Strings,''
Nucl.\ Phys.\ B {\bf 330}, 103 (1990)\semi
K.~A.~Intriligator,
``Bonus Symmetry In Conformal Field Theory,''
Nucl.\ Phys.\ B {\bf 332}, 541 (1990).
}

\lref\LustWR{
D.~L\"ust,
``Intersecting Brane Worlds And Their Effective Interactions,''
Prog.\ Theor.\ Phys.\ Suppl.\  {\bf 152}, 59 (2004).
}

\lref\brunnew{I. Brunner, K. Hori, K. Hosomichi and J. Walcher,
``Orientifolds of Gepner Models,''
hep-th/0401137.
}

\lref\BlumenhagenSU{
R.~Blumenhagen,
``Supersymmetric orientifolds of Gepner models,''
JHEP {\bf 0311}, 055 (2003)
[arXiv:hep-th/0310244].
}

\lref\GepnerQI{
D.~Gepner,
``Space-Time Supersymmetry In Compactified String Theory And Superconformal Models,''
Nucl.\ Phys.\ B {\bf 296}, 757 (1988).
}

\lref\UrangaPZ{
A.~M.~Uranga,
``Chiral four-dimensional string compactifications with intersecting  D-branes,''
Class.\ Quant.\ Grav.\  {\bf 20}, S373 (2003)
[arXiv:hep-th/0301032].
}

\lref\RecknagelSB{
A.~Recknagel and V.~Schomerus,
``D-branes in Gepner models,''
Nucl.\ Phys.\ B {\bf 531}, 185 (1998)
[arXiv:hep-th/9712186].
}

\lref\brunschom{
I.~Brunner and V.~Schomerus,
``D-branes at singular curves of Calabi-Yau compactifications,''
JHEP {\bf 0004}, 020 (2000)
[arXiv:hep-th/0001132].
}

\lref\fix{
J.~Fuchs, P.~Kaste, W.~Lerche, C.A.~L\"utken, C.~Schweigert and
J.~Walcher,
``Boundary Fixed Points, Enhanced Gauge Symmetry and Singular Bundles
on K3,''Nucl. Phys.B {\bf 598}, 57 - 72 (2001),
[arXiv:hep-th/0007145].
}

\lref\BrunnerJQ{
I.~Brunner, M.~R.~Douglas, A.~E.~Lawrence and C.~R\"omelsberger,
``D-branes on the quintic,''
JHEP {\bf 0008}, 015 (2000)
[arXiv:hep-th/9906200].
}

\lref\BrunnerZM{
I.~Brunner and K.~Hori,
``Orientifolds and mirror symmetry,''
arXiv:hep-th/0303135.
}

\lref\BrunnerEM{
I.~Brunner and K.~Hori,
``Notes on orientifolds of rational conformal field theories,''
arXiv:hep-th/0208141.
}

\lref\BrunnerFS{
I.~Brunner,
``On orientifolds of WZW models and their relation to geometry,''
JHEP {\bf 0201}, 007 (2002)
[arXiv:hep-th/0110219].
}

\lref\HuiszoonAI{
L.~R.~Huiszoon and K.~Schalm,
``BPS orientifold planes from crosscap states in Calabi-Yau  compactifications,''
JHEP {\bf 0311}, 019 (2003) [arXiv:hep-th/0306091].
}

\lref\HuiszoonVY{
L.~R.~Huiszoon,
``Comments on the classification of orientifolds,''
Class.\ Quant.\ Grav.\  {\bf 20}, S509 (2003)
[arXiv:hep-th/0212244].
}

\lref\FuchsCM{
J.~Fuchs, L.~R.~Huiszoon, A.~N.~Schellekens, C.~Schweigert and J.~Walcher,
``Boundaries, crosscaps and simple currents,''
Phys.\ Lett.\ B {\bf 495}, 427 (2000)
[arXiv:hep-th/0007174].
}

\lref\HuiszoonGE{
L.~R.~Huiszoon and A.~N.~Schellekens,
``Crosscaps, boundaries and T-duality,''
Nucl.\ Phys.\ B {\bf 584}, 705 (2000)
[arXiv:hep-th/0004100].
}

\lref\HuiszoonXQ{
L.~R.~Huiszoon, A.~N.~Schellekens and N.~Sousa,
``Klein bottles and simple currents,''
Phys.\ Lett.\ B {\bf 470}, 95 (1999)
[arXiv:hep-th/9909114].
}

\lref\GovindarajanVP{
S.~Govindarajan and J.~Majumder,
``Crosscaps in Gepner models and type IIA orientifolds,''
JHEP {\bf 0402}, 026 (2004)
[arXiv:hep-th/0306257].
}

\lref\GovindarajanVV{
S.~Govindarajan and J.~Majumder,
``Orientifolds of type IIA strings on Calabi-Yau manifolds,''
arXiv:hep-th/0305108.
}

\lref\BlumenhagenTJ{
R.~Blumenhagen and A.~Wisskirchen,
``Spectra of 4D, N = 1 type I string vacua on non-toroidal CY threefolds,''
Phys.\ Lett.\ B {\bf 438}, 52 (1998)
[arXiv:hep-th/9806131].
}

\lref\AngelantonjMW{
C.~Angelantonj, M.~Bianchi, G.~Pradisi, A.~Sagnotti and Y.~S.~Stanev,
``Comments on Gepner models and type I vacua in string theory,''
Phys.\ Lett.\ B {\bf 387}, 743 (1996)
[arXiv:hep-th/9607229].
}

\lref\AldazabalUB{
G.~Aldazabal, E.~C.~Andres, M.~Leston and C.~Nunez,
``Type IIB orientifolds on Gepner points,''
JHEP {\bf 0309}, 067 (2003)
[arXiv:hep-th/0307183].
}

\lref\FuchsGV{
J.~Fuchs, C.~Schweigert and J.~Walcher,
``Projections in string theory and boundary states for Gepner models,''
Nucl.\ Phys.\ B {\bf 588}, 110 (2000)
[arXiv:hep-th/0003298].
}

\lref\rbgklnon{R.~Blumenhagen, L.~G\"orlich, B.~K\"ors and D.~L\"ust,
``Noncommutative Compactifications of Type I Strings on Tori with Magnetic
Background Flux'', JHEP {\bf 0010}, 006 (2000), [arXiv:hep-th/0007024] \semi
R.~Blumenhagen, B.~K\"ors and D.~L\"ust,
``Type I Strings with $F$ and $B$-Flux'', JHEP {\bf 0102}, 030 (2001),
[arXiv:hep-th/0012156].
}

\lref\raads{C.~Angelantonj, I.~Antoniadis, E.~Dudas, A.~Sagnotti, ``Type I
Strings on Magnetized Orbifolds and Brane Transmutation'',
Phys. Lett. B {\bf 489}, 223 (2000), [arXiv:hep-th/0007090].
}

\lref\rbbkl{R.~Blumenhagen, V.~Braun, B.~K\"ors and D.~L\"ust,
``Orientifolds of K3 and Calabi-Yau Manifolds with Intersecting D-branes'',
JHEP {\bf 0207}, 026  (2002), [arXiv:hep-th/0206038].
}

\lref\rbtw{R.~Blumenhagen, T.~Weigand, ``Chiral Supersymmetric Gepner
Model Orientifolds'', JHEP {\bf 0402}, 041 (2004), [arXiv:hep-th/0401148].
}

\lref\sagn{A.~Sagnotti, Y.~Stanev, ``Open Descendants in Conformal
Field Theory'', Fortsch.Phys. {\bf 44}, 585- 596 (1996), [arXiv:hep-th/9605042].
}

\lref\rcveticb{M.~Cvetic, G.~Shiu and  A.M.~Uranga,  
``Chiral Four-Dimensional N=1 Supersymmetric Type IIA Orientifolds from
Intersecting D6-Branes'', Nucl. Phys. B {\bf 615}, 3  (2001), [arXiv:hep-th/0107166]\semi
R.~Blumenhagen, L.~G\"orlich and T.~Ott,
``Supersymmetric intersecting branes on the type IIA 
$T^6/Z(4)$  orientifold'', JHEP {\bf 0301}, 021 (2003),
[arXiv:hep-th/0211059]\semi
G.~Honecker,
``Chiral supersymmetric models on an orientifold of Z(4) x Z(2) with 
 intersecting D6-branes,''
Nucl.\ Phys.\ B {\bf 666}, 175 (2003)
[arXiv:hep-th/0303015]\semi
M.~Larosa and G.~Pradisi,
``Magnetized four-dimensional Z(2) x Z(2) orientifolds,''
Nucl.\ Phys.\ B {\bf 667}, 261 (2003)
[arXiv:hep-th/0305224]\semi
M.~Cvetic and I.~Papadimitriou,
 ``More supersymmetric standard-like models from intersecting D6-branes on
 type IIA orientifolds,''
Phys.\ Rev.\ D {\bf 67}, 126006 (2003)
[arXiv:hep-th/0303197]\semi
M.~Cvetic, T.~Li and T.~Liu,
``Supersymmetric Pati-Salam models from intersecting D6-branes: A road to the
 standard model,''
arXiv:hep-th/0403061.
}

\lref\rbachas{C.~Bachas, ``A Way to Break Supersymmetry'', [arXiv:hep-th/9503030]\semi 
M.~Berkooz, M.R.~Douglas and R.G.~Leigh, ``Branes Intersecting
at Angles'', Nucl. Phys. B {\bf 480}, 265  (1996), [arXiv:hep-th/9606139].
}

\lref\rafiru{G.~Aldazabal, S.~Franco, L.E.~Ibanez, R.~Rabadan, A.M.~Uranga,
`` $D=4$ Chiral String Compactifications from Intersecting Branes'',
J.\ Math.\ Phys.\  {\bf 42}, 3103 (2001), [arXiv:hep-th/0011073]\semi
G.~Aldazabal, S.~Franco, L.E.~Ibanez, R.~Rabadan, A.M.~Uranga,
``Intersecting Brane Worlds'', JHEP {\bf 0102}, 047 (2001), [arXiv:hep-ph/0011132].
}

\lref\DijkstraYM{
T.~P.~T.~Dijkstra, L.~R.~Huiszoon and A.~N.~Schellekens,
 ``Chiral Supersymmetric Standard Model Spectra from Orientifolds of Gepner
 Models,''
arXiv:hep-th/0403196.
}

\lref\DouglasUM{
M.~R.~Douglas,
``The statistics of string / M theory vacua,''
JHEP {\bf 0305}, 046 (2003)
[arXiv:hep-th/0303194].
}

\lref\BianchiYU{
M.~Bianchi and A.~Sagnotti,
``On The Systematics Of Open String Theories,''
Phys.\ Lett.\ B {\bf 247}, 517 (1990)\semi
M.~Bianchi and A.~Sagnotti,
``Twist Symmetry And Open String Wilson Lines,''
Nucl.\ Phys.\ B {\bf 361}, 519 (1991)\semi
M.~Bianchi, G.~Pradisi and A.~Sagnotti,
``Toroidal compactification and symmetry breaking in open string theories,''
Nucl.\ Phys.\ B {\bf 376}, 365 (1992).
}

\lref\Aldaneu{
G.~Aldazabal, E.~C.~Andres, J.~E.~Juknevich
``Particle Models form Orientifolds at Gepner-orbifold Points,''
arXiv:hep-th/0403262.
}

\lref\PradisiQY{
G.~Pradisi, A.~Sagnotti and Y.~S.~Stanev,
``Planar duality in SU(2) WZW models,''
Phys.\ Lett.\ B {\bf 354}, 279 (1995)
[arXiv:hep-th/9503207]\semi
G.~Pradisi, A.~Sagnotti and Y.~S.~Stanev,
``The Open descendants of nondiagonal SU(2) WZW models,''
Phys.\ Lett.\ B {\bf 356}, 230 (1995)
[arXiv:hep-th/9506014]\semi
G.~Pradisi, A.~Sagnotti and Y.~S.~Stanev,
``Completeness Conditions for Boundary Operators in 2D Conformal Field Theory,''
Phys.\ Lett.\ B {\bf 381}, 97 (1996)
[arXiv:hep-th/9603097].
}

\Title{\vbox{
 \hbox{DAMTP-2004-36}
 \hbox{hep-th/0403299}}}
{\vbox{\centerline{ A Note on Partition Functions of} 
\vskip 0.3cm \centerline{Gepner Model Orientifolds}
}}
\centerline{Ralph Blumenhagen{$^1$}, Timo Weigand{$^2$} }
\bigskip\medskip
\centerline{ {\it DAMTP, Centre for Mathematical Sciences,}}
\centerline{\it Wilberforce Road, Cambridge CB3 0WA, UK}
\centerline{\tt Email:\vtop{{\hbox{$^1$:R.Blumenhagen@damtp.cam.ac.uk}}
                            {\hbox{$^2$:T.Weigand@damtp.cam.ac.uk}}} }
\bigskip
\bigskip

\centerline{\bf Abstract}
\noindent
In this note we generalize the description of simple
current extended Gepner Model 
orientifolds as presented  in hep-th/0401148 to the case of even 
levels and non-trivial dressings of the parity transformation.
We provide a comprehensive list of all the important ingredients for the
construction of such orientifolds. Namely we present
explicit expressions for the Klein-bottle, annulus
and M\"obius strip amplitudes and derive the general
tadpole cancellation conditions.   
As an example we construct  a supersymmetric  Pati-Salam 
like model. 


\Date{03/2004}
\newsec{Introduction}

It is still an open questions whether string theory really contains
solutions to its equations of motion resembling the particle
physics we observe at low energies. In order to definitely answer this question,
we eventually have no other choice than to construct various string backgrounds
and check whether Standard Model like features can be achieved. 
Various classes of four-dimensional string compactifications have been 
studied in some detail during the last twenty years.

Most recently, there has been extended work on the construction of models using
intersecting D-branes as an essential ingredient to get unitary gauge symmetries 
and chirality \refs{\rbachas\rbgklnon\raads\rafiru\rcveticb\UrangaPZ-\LustWR}. 
Though non-supersymmetric Standard-like models could be found 
fairly generically \refs{\rafiru}, an intersecting brane realization of the MSSM using just toroidal
orbifold backgrounds turned out to be much harder to achieve 
\refs{\rcveticb}. 

After some earlier studies \refs{\AngelantonjMW,\BlumenhagenTJ}, 
during the last months we have seen a renewed interest in the construction of orientifolds
of Gepner models \refs{\AldazabalUB\BlumenhagenSU\brunnew\rbtw\DijkstraYM-\Aldaneu}, 
which allows one to really move beyond the framework
of toroidal orbifolds and to study intersecting brane worlds on small
scale Calabi-Yau manifolds \rbbkl.
Historically, essentially two approaches have been followed so far. 
The first one starts on the level of one-loop partition functions and extracts the tadpoles
from the explicitly computed Klein-bottle, annulus and M\"obius strip  amplitudes 
\refs{\AngelantonjMW,\BlumenhagenTJ,\AldazabalUB,\BlumenhagenSU,\rbtw,\Aldaneu}. 
The second approach starts directly on the level
of crosscap states in these conformal field theories  
\refs{\BianchiYU\PradisiQY\HuiszoonXQ\HuiszoonGE\FuchsCM\BrunnerFS
\HuiszoonVY\BrunnerEM\BrunnerZM\GovindarajanVV\HuiszoonAI-\GovindarajanVP,\brunnew,
\DijkstraYM}
and then introduces boundary states to cancel the crosscap tadpoles.  
From there one moves forward to determine the loop-channel amplitudes. 
Apparently, these two approaches are  completely equivalent. 

The aim of this note is nothing more than to bring both approaches on equal footing and to 
relax the assumptions under which the results
of \refs{\BlumenhagenSU, \rbtw} have been derived using the first approach. 
More concretely, we generalize the one-loop partition functions,
as derived in   \refs{\BlumenhagenSU, \rbtw} for levels being odd, 
to the case of even levels. Moreover,  on the level of partition
functions we implement additional dressings of the word-sheet parity symmetry 
and identify them with the dressings introduced  
in \brunnew\ in the crosscap state approach.
As expected, all the physical information can be
read off entirely from the various  amplitudes.
We will end up with a collection  of very explicit and general 
one-loop partition functions and tadpole
cancellation conditions covering simple current extensions
of all  168 Gepner models with additional dressings of the parity symmetry.
In fact providing a  compact collection of the main relevant formulas
for constructing supersymmetric Gepner Model orientifolds was one of the  motivations
for writing this letter. We hope that these expressions
turn out to be useful  for a systematic  search for standard-like models respectively 
for providing a statistical ensemble in the spirit of \DouglasUM.

This paper is organized as follows. In section two we generalize the
computation of the A-type loop channel Klein-bottle amplitude to the case
of even levels allowing as well certain  dressings of the parity symmetry. 
After deriving the tree-channel amplitude using the methods
of  \refs{\BlumenhagenSU, \rbtw}, we determine explicitly the NS-NS sector crosscap
state including all sign factors. Section three deals with the open
string sector and after computing the M\"obius strip amplitude we fix
the action of the dressed parity transformation on the boundary states. 
In section 4 we derive the form of the tadpole cancellation conditions
and present a Pati-Salam like  model in section 5, providing evidence
that phenomenologically interesting Standard-like models are likely to
be contained in the huge set of Gepner Model orientifolds \DijkstraYM.

\newsec{Orientifolds of extended Gepner models: The A-type Klein bottle}

In \rbtw\ we have derived one-loop partition functions for 
simple current \SchellekensAM\
extended  Gepner model \GepnerQI\ orientifolds under the assumption of all levels
being odd.
In this section we repeat the analysis but give up this latter restriction and allow
some of the levels to be even. 
In this case some of the 168 Gepner models with $c=9$ have only four tensor factors, but,
as pointed out in \brunnew, this case should be treated
as having five tensor factors with $k_5=0$. 

For an explanation of the notation to be  used in the following  and an introduction into Gepner model 
orientifolds we would like to
refer the reader to our former papers \refs{\BlumenhagenSU,\rbtw} and
references therein.
Our starting point here is the simple current extended charge conjugated
Gepner model torus partition function
\eqn\gepo{\eqalign{
Z_C (\tau, \bar{\tau})  &= {1\over N} {1\over 2^r} 
{( {\rm Im} \tau)^{ - 2}\over | \eta (q)|^2 }
 \sum_{b_0 = 0}^{K-1}
\sum_{b_1, \ldots, b_r = 0 }^{1} 
\sum_{\tau_1= 0}^{{\cal N}_1 -1} \ldots \sum_{\tau_I= 0}^{{\cal N}_I -1}
{\sum_{\lambda,\mu}}^{\beta}\,
( -1)^{s_0} \cr
&\prod_{\alpha= 1}^I \, \delta^{(1)} \left( Q^{(\alpha)}_{\lambda, - \mu}  + 
2 \,\tau_{\alpha} \hat{Q}^{(\alpha)}(J_{\alpha}) \right)
\, \chi^{\lambda}_{\mu}(q) \, \,  
\chi^{\lambda}_{-\mu + b_0 \beta_0 + b_1 \beta_1 + \ldots + b_r \beta_r + 
\sum_{\alpha} 2\, \tau_{\alpha} j_{\alpha}  }(\bar{q}) ,}}
with $K={\rm lcm}(4,2k_j+4)$ and where we have taken $I$ different mutually local 
simple currents $J_\alpha$ of length
${\cal N}_\alpha$ and where $Q^{(\alpha)}_{\lambda,\mu}$ denotes the monodromy
charge of the field $(\lambda,\mu)$ with respect to the simple current $J_\alpha$.
Let us assume that the simple currents only act on the internal sector, so that
they can be brought to the form
\eqn\gepd{
j_{\alpha} = ( 0; m_1^{\alpha}, \ldots, m_r^{\alpha}; 0, \ldots, 0) }
with all $m_j^{\alpha}$ even.
As is evident from \gepo, the states surviving the $\Omega$-projection in the charge conjugated
(A-type) partition function have to satisfy
\eqn\kaa{
\mu \cong  -\mu + b_0 \beta_0 + b_1 \beta_1 + \ldots + b_r \beta_r + \sum_{\alpha} 
\tau_{\alpha} j_{\alpha},    }
i.e.
\eqn\kab{\eqalign{
m_j &= b + \sum_{\alpha} \tau_{\alpha}\frac{ m_j ^{\alpha}}{2}  + \frac{1}{2}\eta_j (k_j +2) \quad
{\rm mod} \ (k_j +2)  \quad {\rm for\ all}\  j,  \cr
   s_0 &= b + \sum_i b_i   \quad {\rm mod} \ 2,  \cr
   s_j & = b + b_j + \eta_j\quad{\rm mod} \, 2 }}
for some $b$ in the range $\{0, \ldots, \frac{K}{2} -1 \}$, $b_j = 0,1$.
The only change compared to the case of only odd levels is the
appearance of  $\eta_j$, which  takes the values
$\eta_j =0,1$ in every tensor factor where $l_j =\frac{ k_j}{2}$ and 
vanishes otherwise. Therefore it is only  present for even $K' = {\rm lcm}(k_j +2)$.
The origin of $\eta_j$ is due to the fact that for even levels 
the value $l_j =\frac{ k_j}{2}$ is invariant under the reflection symmetry
$(l_j,m_j,s_j)\to (k_j-l_j,m_j+k_j+2,s_j+2)$, thus leading to the existence
of shorter simple current orbits.
The constraints on $s_j$ and $s_0$ imply 
\eqn\kabd{\sum_j \eta_j = 0 \quad{\rm mod} \, 2. }
Since our aim is to exploit  the resulting expressions for a
systematic examination of the spectrum, it turns out to be useful to
require that for all pairs of simple currents 
$Q^{(\alpha)}(J_{\beta})=\sum_j (m_j^\alpha  m_j^\beta)/ (2k_j+4)$,
is an even integer. 
This will simplify the calculations and the resulting expressions considerably.
These projections are then implemented as in \rbtw.
As is well known, however, the orientifold projection is by no means
unique in the sense that one is always free to dress the characters
which survive the projection with additional signs consistent with the
fusion rules \sagn. 

In view of the free parameters in \kaa\ and the various relations \kab\
between them,  we 
define the orientifold projection  $\Omega_{{\Delta}_j, \omega,
\omega_\alpha}$ by including the sign factors
\eqn\dressing{ (-1)^{ {\omega} \, (b +s_0) + \sum_j  {\Delta}_j \eta_j + 
 \sum_\alpha  {\omega}_\alpha\, \tau_\alpha }}
for $\Delta_i, \omega, \omega_\alpha= 0,1$.
Note  that the $\Delta_j$ only
have a non-trivial effect if $k_j$ is even. Moreover, the combination $(b +s_0)$
is just right for the $\omega$ dressing to preserve supersymmetry
of the resulting Klein-bottle amplitude and is only well defined for
$K'$ even. Similarly, a non-trivial
simple current dressing, $\omega_\alpha=1$,  is only allowed for ${\cal N}_\alpha$ even. 
Independently of these optional parity
dressings, consistency with  our results from \rbtw\  for the case of all 
levels being odd requires a factor of $\prod_{k<l} (-1)^{\eta_k
\eta_l}$.
Then, the overall A-type Klein bottle can be written as 
\eqn\kaj{\eqalign{
K^{A}({\Delta}_j, \omega, \omega_\alpha) &= 4 \int_0^{\infty} \frac{dt}{t^3} \frac{1}{2^{r+1}} 
  \frac{1}{\eta( 2it)^2}  {\sum_{\lambda,\mu}}^{\beta}\,
\sum_{\eta_1,\ldots,\eta_r=0}^1 
 \sum_{b=0}^{\frac{K}{2}-1} \,  
\sum_{\tau_1= 0}^{{\cal N}_1 -1} \ldots \sum_{\tau_I= 0}^{{\cal N}_I -1} \,\,
(-1)^{s_0}\, (-1)^{\omega  (b +s_0)} \cr
& (-1)^{ \sum_j  {\Delta}_j \eta_j} \,  
     (-1)^{ \sum_\alpha  {\omega}_\alpha \tau _\alpha} \,  \,
   \delta^{(2)}_{\sum_j \eta_j, 0} \,\, \left( \prod_{k<l} (-1)^{\eta_k \eta_l} \right)
  \left( \prod_j \delta_{l_j \eta_j, \frac{k_j}{2} \eta_j} \right) \cr
& \left( \prod_{\alpha} \delta^{(1)}_{ \sum_j {1\over 4} \eta_j
m_j^\alpha,0} \right)\,
 \left( \prod_{j=1}^r \, \delta^{(k_j +2)}_{m_j, b + \sum_{\alpha}
{1\over 2}\tau_{\alpha} m_j^{\alpha} + \eta_j \frac{1}{2}(k_j+2)}\right)  \,
 \chi^{\lambda}_{\mu} ( 2 i t ), }}
where the first term in the last line is a remnant of the monodromy 
charge constraint in \gepo. The  tree-channel amplitude is modified accordingly as
\eqn\kak{\eqalign{
\widetilde{K}^{A}({\Delta}_j, \omega,\omega_\alpha)& =  
 \frac{2^4 \prod_\alpha {\cal N}_{\alpha}}{ 2^{\frac{3r}{2}} 
 \prod_j \sqrt{ k_j + 2}} 
\int_0^{\infty} dl \frac{1}{\eta^2(2il)} 
{\sum_{\lambda',\mu'}}^{ev}\,
\sum_{\eta_1, \ldots, \eta_r=0}^{1} \,
\sum_{\nu_0 = 0}^{K-1} \,
\sum_{\nu_1, \ldots, \nu_r=0}^{1} \,
\sum_{\epsilon_1, \ldots, \epsilon_r=0}^{1}  \cr 
&\left( \prod_{k<l} (-1)^{\eta_k \eta_l} \right)\,
\left( \prod_{\a} \delta^{(1)}_{\sum_j {1\over 4}\eta_j m_j^\alpha,0} \right)\,
\left( \prod_{\alpha} \delta^{(2)}_{Q^{(\alpha)}_{\lambda',
\mu' + (1- \vec{\epsilon})(\vec{k} +2)},\omega_\alpha } \right) \cr
&\delta_{\sum_j \eta_j, 0}^{(2)}\,\,
\delta^{(4)}_{s'_0 + \nu_0 + 2 \sum \nu_j +2, 2\omega }\,\,
\delta^{(2)}_{ \sum_j \frac{1}{k_j +2} (m'_j  + (1-
\epsilon_j)(k_j + 2)), \omega} \cr  
&\prod_{j=1}^r  \Biggl( \frac{ P_{l_j',\epsilon_j k_j} P_{l_j', (\epsilon_j +
\eta_j) k_j}}{S_{l_j',0}} \,\, \delta^{(2)}_{\eta_j k_j,0} \,\, 
(-1)^{\eta_j   \left( \frac{m_j'}{2} + \nu_0 + \Delta_j +(1-\epsilon_j) \right)}   \cr
&  \delta^{(2)}_{m'_j + (1- \epsilon_j)(k_j + 2),0 } \,\,
\delta^{(4)}_{s'_j+  \nu_0 +2\nu_j + 2(1-\epsilon_j),0} \Biggr)\,\,
\chi^{\lambda'}_{\mu'}(2il), }}
where we have introduced the short-hand notation $(1- \vec{\epsilon})(\vec{k} +2)$
for the vector 
$\mu_{\vec\epsilon}=(0;(1-\epsilon_1)(k_1+2),\ldots,(1-\epsilon_5)(k_5+2);0,\ldots,0)$.
Note that besides the appearance of the sum over the parameters $\eta_j$ 
also the conditions on the monodromy charges with respect to the additional
simple currents changes slightly as compared to the case of all levels being odd.
 
From the tree-channel Klein bottle, we can read off the crosscap state
up to overall signs and complex phases which cancel  in the
overlap. These phases fall into two classes: Those depending only on the
states contributing to the crosscap and those which are a function of the
parameters of the dressings. For the determination of the signs in the first class
we follow the method presented in \refs{\BlumenhagenSU,\rbtw} (which
was shown  to work in the NS-NS sector and is therefore sufficient for
supersymmetric models).
The second  class of phases has no physical meaning as they can be rotated
away.
Once a particular choice is made, however, it determines the parity
action on the boundary states uniquely, as we will see from the
M\"obius amplitude. 

For pure convenience, we choose to include the phase factor
\eqn\phasb{   (-1)^{\omega
\frac{s'_0}{2}} e^{i \pi \sum_j  \frac{  \Delta_j\, (m_j'+k_j+2)}{k_j +2}}}
into the crosscap state. Independently of this convention, an additional ${\rm exp}({i \pi \sum_j
{\Delta_j}\epsilon_j})$ is really required to obtain \kak\ correctly, so that 
the final crosscap state takes the form
\eqn\kam{\eqalign{
\big| C; {\Delta}_j,\omega, \omega_\alpha \big>_{NS} &= \frac{1}{\kappa^A_c}
{\sum_{\lambda',\mu'}}^{ev}\, 
\sum_{\nu_0 = 0}^{{K\over 2}-1} \,
\sum_{\nu_1, \ldots, \nu_r=0}^{1} \,
\sum_{\epsilon_1, \ldots, \epsilon_r=0}^{1}\  
(-1)^{\nu_0} \left( \prod_{k<l} (-1)^{\nu_k \nu_l}\right) \,
(-1)^{\sum_j \nu_j} \cr
& (-1)^{\omega \frac{s'_0}{2}}\, 
e^{i \pi \sum_j \frac{\Delta_j}{k_j +2} (m_j' + (1-\epsilon_j)(k_j
 +2))}  \, \left( \prod_\alpha \delta^{(2)}_{Q^{(\alpha)}_{\lambda',
\mu' + (1- \vec{\epsilon})(\vec{k} +2)},\omega_\alpha } \right)  \cr
&  \delta^{(4)}_{s'_0 + 2 \nu_0 + 2 \sum \nu_j +2, 2 \omega }\,
\delta^{(2)}_{ \sum_j \frac{1}{k_j +2} (m'_j  + (1-
\epsilon_j)(k_j + 2)), \omega } \prod_{j=1}^r  \Biggl( \sigma(l_j', m_j', s_j')  
\frac{P_{l'_j, \epsilon_j \, k_j}}{\sqrt{S_{l'_j,0}}}  \cr
& (-1)^{\epsilon_j \frac{m_j' + s_j'}{2}}\, \delta^{(2)}_{m'_j + (1- \epsilon_j)(k_j + 2),0 } \,
 \delta^{(4)}_{s'_j+ 2\nu_0 +2\nu_j + 2(1-\epsilon_j),0} \Biggr)\,\,
\big|{\lambda'},{\mu'}\big>\big>_c,}}
where 
\eqn\kah{
\left(\frac{1}{\kappa_c^A}\right)^2 = 
\frac{ 2^5 \big( \prod_{\alpha=1}^I {\cal N}_{\alpha}\big) }{ 2^{\frac{3r}{2}}K
 \prod_j \sqrt{k_j + 2} }. }
Comparing this crosscap state for  $\omega_\alpha=0$ 
to the one used in \brunnew\ one finds complete agreement.
Therefore we conclude that the $\Delta_j$ really define the various phase dressings and $\omega$ 
the quantum dressing of the parity transformation as introduced in \brunnew.  
Note that the crosscap state 
\kam\ in addition  includes the  non-trivial simple current dressings $\omega_\alpha$.
 
As anticipated before, the form of the crosscap state for the case of even levels does not differ
at all from its analogue for the case of all levels being odd. The new parameters, $\eta_j$,
in the former case simply arise from additional contributions in 
the overlap of the crosscap state with itself and therefore arise automatically from \kam.

\newsec{Open string one loop amplitudes}

As usual, in order to cancel the massless tadpoles of the orientifold planes
one introduces A-type boundary states, which for a  simple current extension
have the form
\eqn\ana{\eqalign{
 \big|a\big>_A = \big|S_0; (L_j, M_j, S_j)_{j=1}^{r}\big>_A &=
\frac{1}{\kappa_{a}^A} 
{\sum_{\lambda',\mu'}}^{\beta}\, 
\prod_{\alpha} \delta^{(1)} \big( Q^{(\alpha)}_{\lambda', \mu'} \big) 
(-1)^{\frac{s'^2_0}{2}} e^{ -i\pi \frac{s'_0 S_0}{2}} \cr
&\prod_{j=1}^r \bigg( \frac{S_{l'_j, L_j}}{\sqrt{S_{l'_j,0}}} \,\,
e^{i \pi \frac{m'_j M_j}{k_j +2} } \, e^{-i \pi \frac{s'_j S_j}{2} }\bigg) 
\big|{\lambda',\mu'}\big>\big> }}
with the normalization
\eqn\and{
\frac{1}{\left(\kappa_{a}^A\right)^2 } = 
\frac {K \,\left(\prod_{\alpha} {\cal N}_{\alpha}\right)}{2^{\frac{r}{2} +1} \prod_j \sqrt{k_j +2} }. }
Note that boundary state labels connected by the action of the simple currents $J_\alpha$ 
describe identical D-branes.
In order to finally read off the massless spectrum, we have to
transform their overlap into loop channel
\eqn\anc{\eqalign{
A_{\tilde{a}\, a}^A &= N_{a} N_{\tilde{a}}\, {1\over 2^{r+1}}\,
\int_0^{\infty} \frac{dt}{t^3} \frac{1}{\eta^2(it)} 
 {\sum_{\lambda,\mu}}^{ev}\,
\sum_{\nu_0 = 0}^{K-1}
\sum_{\nu_1, \ldots, \nu_r = 0 }^{1}
\sum_{\epsilon_1, \ldots,\epsilon_r=0}^1 
\sum_{\sigma_1= 0}^{{\cal N}_1 -1} \ldots \sum_{\sigma_I= 0}^{{\cal N}_I -1}  
( -1)^{\nu_0} \cr
&\delta^{(4)}_{s_0, 2+ \tilde{S}_0 - S_0 - \nu_0 - 2 \sum_j \nu_j} 
\prod_{j=1}^r \bigg( N^{|\epsilon_j k_j - l_j|}_{L_j, \tilde{L}_j}\,\,
\delta ^{(2k_j +4)}_{m_j + M_j - \tilde{M}_j + \nu_0 + \sum_{\alpha} 
\sigma_{\alpha} m_j^{\alpha} + \epsilon_j (k_j+2), 0} \cr
&\delta^{(4)}_{s_j, \tilde{S}_j - S_j - \nu_0 - 2 \nu_j + 2 \epsilon_j} \bigg) 
\chi_{\mu}^{\lambda} (it).  }}
It is well known that for even levels some of the boundary states \ana\ are 
not fundamental and split into fractional branes. These so-called resolved
boundary states have been constructed in \refs{\brunschom, \fix}. Here just for keeping the presentation
simple we work with the unresolved Recknagel/Schomerus \refs{\RecknagelSB,\BrunnerJQ} 
boundary states \ana.

Let us now address the issue  of the action of $\Omega_{{\Delta}_j, \omega, \omega_\alpha}$ on a
boundary state. For this purpose, we compute the overlap of a boundary state
with the crosscap state \kam, which by the way is not different for
the resolved boundary states, as the crosscap state only contains untwisted
contributions.  After transforming into loop channel
we obtain 
\eqn\moe{\eqalign{
M_a^{A, NS}( {\Delta}_j, \omega, \omega_\alpha)   &= (-1)^s N_{a} \, \,
\frac{1}{2^{r+1}}
\int_0^{\infty} \frac{dt}{t^3} \frac{1}{\hat\eta^2(it+\frac{1}{2})} 
{\sum_{\lambda,\mu}}^{ev}\,  \sum_{\nu_0 = 0}^{\frac{K}{2}-1}
\sum_{\epsilon_1, \ldots, \epsilon_r=0}^{1}  
\sum_{\sigma_1 = 0}^{{\cal N}_1 -1} \ldots \sum_{\sigma_{I} =
0}^{{\cal N}_{I} -1}   \cr
&(-1)^{\omega (\nu_o+ \frac{s_0}{2})}\,\, (-1)^{\sum_\alpha \omega_\alpha \tau_\alpha}\,\,
\left( \prod_{k<l} (-1)^{\rho_k \rho_l} \right)\, 
\delta^{(2)}_{\sum_j \rho_j, 0}\, \, 
\delta^{(2)}_{s_0,0} \cr
&\prod_{j=1}^r  \Biggl( \sigma_{(l_j, m_j, s_j)}\, Y^{l_j}_{L_j,
\epsilon_j k_j} \, \,
\delta^{(2)}_{s_j,0}  \,\,
\delta^{(2k_j +4)}_{2 (M_j- \Delta_j)+ m_j + 2 \nu_0+ \sum_{\alpha} \sigma_{\alpha} m_j^{\alpha} + 
\epsilon_j (k_j+2), 0} \cr
& 
(-1)^{\frac{\epsilon_j}{2} [2S_j  -s_j- 2\epsilon_j]}
(-1)^{\frac{(1-\epsilon_j)}{2} [2M_j  -m_j+ \epsilon_j (k_j+2)]} \Biggr)\,\
\hat{\chi}^{\lambda}_{\mu} (it+\frac{1}{2}),     }}
where
\eqn\rhoo{\eqalign{  r &= 4 s +1, \cr
           \rho_j&={s_0+s_j\over 2} + \omega+ \epsilon_j-1, }}
and the $Y$-tensor is defined as
\eqn\ymatr{  Y_{l_1,l_2}^{l_3}=\sum_{l=0}^k  {S_{l_1,l}\, P_{l_2,l} \, P_{l_3,l}\over
                               S_{0,l} } .} 
Requiring that the M\"obius amplitude \moe\ is consistent with the annulus amplitude \anc\
for a D-brane and its $\Omega_{{\Delta}_j, \omega,\omega_\alpha}$ image, we can derive
 the action of $\Omega_{{\Delta}_j, \omega, \omega_\alpha}$ on a boundary state.
First note that $\Omega$ itself reverses the sign of the labels $S_0, M_j, S_j$. 
The phase dressings shift the $M_j$ to $M_j+2\Delta_j$ and  the
$\omega$ dressing changes the GSO projection in \moe\ and therefore
maps a brane to its anti-brane, which can also be described by the shift $S_0\to S_0+2$.
Finally, the $\omega_\alpha$ dressings only change some sign factors
in \moe\ and therefore should leave a boundary state invariant. To summarize, the 
entire action of  $\Omega_{{\Delta}_j, \omega,\omega_\alpha}$ on a boundary state
is given by
\eqn\acti{ \big| S_0, \prod_j (L_j, M_j, S_j)\big>\big> \rightarrow \big| -S_0 +2\omega,
 \prod_j (L_j,- M_j +2\Delta_j,-S_j)\big>\big> . }

In particular, the invariant branes of the pure non-extended Gepner Model 
are now classified by
\eqn\bb{
|S_0, \prod_{j=1}^{r'}(\frac{k_j}{2}, \Delta_j + \frac{k_j
 +2}{2}, S_j) \prod_{j=r'+1}^r (L_j, \Delta_j, S_j) \big>\big> }
for $(r'-\omega)$ even and the $M_j$ chosen modulo $(k_j+2)$.
A boundary state is supersymmetric relative  to the crosscap state if
\eqn\susy{  {S_0-\omega\over 2} -\sum_j {M_j-\Delta_j \over k_j+2} + 
          \sum_j {S_j \over 2} =0\ {\rm mod}\ 2 .}
From these latter expressions it is clear that the phase dressings can be thought
of as a rotation in the $M_j$ planes, whereas the quantum $\omega$ dressing similarly
can be considered as a rotation in $S_0$ plane. Therefore, from the conformal field
theory point of view the phase shifts and the quantum dressing are completely analogous.   

\vskip 1cm

\newsec{Tadpole cancellation conditions}

The tadpole cancellation conditions contain both the contribution
from the D-branes and from the orientifold planes and
take the general form Tad$_{D}(\lambda,\mu)-4$Tad$_{O}(\lambda,\mu)=0$ for 
the massless fields
$(2)(0,0,0)^5$ and $(0)\prod_j(l_j, l_j, 0)$ with $\sum_j \frac{l_j}{k_j +2}
= 1$.
Up to the  common  factor
\eqn\common{ {\rm const}. 
\times {e^{i \pi \sum_j \frac{\Delta_j}{k_j +2} m_j }\over  \prod_j \sqrt{ S_{l_j,0}} }, }
containing in particular the phase convention mentioned in the end of section 2,
the NS-NS tadpoles of the orientifold plane read 
\eqn\kal{\eqalign{
{\rm Tad}_{O}(\lambda,\mu) &= (-1)^{(1+ \frac{s_0}{2})(1+\omega)} \sum_{\epsilon_1, \ldots,
\epsilon_r = 0}^1 \, \, 
e^{i \pi \sum_j \frac{\Delta_j}{k_j +2} (1-\epsilon_j)(k_j
 +2)}\,\, \left( \prod_{k<l} (-1)^{\epsilon_k \epsilon_l}\right) \cr
&\delta^{(2)}_{\sum \epsilon_j, \omega + \frac{s_0}{2}} \, \, 
 \left( \prod_{\a}
\delta^{(2)}_{Q^{(\alpha)}_{\lambda,
\mu + (1- \vec{\epsilon})(\vec{k} +2)},\omega_\alpha} \right)\, \,  
\prod_j \Biggl(  {\rm sin}\left[\frac{1}{2}(l_j, \epsilon_j k_j)\right]\, \, 
\delta^{(2)}_{l_j+(1-\epsilon_j) k_j,0}\cr
&\delta^{(2)}_{m_j + (1-\epsilon_j)(k_j+2), 0 } \,\,(-1)^{\epsilon_j
 \frac{m_j}{2} } \Biggr). }}
Note that for $k_j$ even only those massless states with $m_j$ even
do have a non-vanishing tadpole on the orientifold plane.  
Collecting all terms from the boundary states and their 
$\Omega_{{\Delta}_j,\omega,\omega_\alpha}$
images, their massless tadpoles read
\eqn\bc{\eqalign{
{\rm Tad}_{D}(\lambda,\mu)& =  
\left(  \prod_{\a}
\delta^{(1)}_{Q^{(\a)}_{\lambda, \mu}}\right)\,
 \sum_{a=1}^N 2\, N_{a}\, {\rm cos}\left[\pi
     \sum_j \frac{m_j (M_j^a-\Delta_j)}{k_j+2} \right] \,
\prod_j {\rm sin}(l_j,L_j^a). }}
By now we have provided a comprehensive  collection of the salient formulas 
needed to construct orientifolds of Gepner Models. 
We featured all one-loop partition 
functions and the resulting tadpole
cancellation conditions  covering simple current extended Gepner model orientifolds with
generally dressed $\Omega_{{\Delta}_j, \omega,\omega_\alpha}$ parity. We hope that these very explicit 
expressions will be  helpful for future work on classifying  semi-realistic models
respectively on carrying out a statistical analysis in the spirit of \DouglasUM.
As a simple example showing that semi-realistic models are possible to get we present
in the final section a two  generation supersymmetric Pati-Salam model. 

\newsec{A Pati-Salam like  example}

We take the $(6)^4$ Gepner model, which 
has Hodge numbers $(h_{21},h_{11})=(1,149)$ but leads
after extending it by the two simple currents
\eqn\exta{  J_1=(0;2,-2,0,0, 0;0,0,0,0,0), \quad\quad J_2=(0;2,2,-4,0,0;0,0,0,0,0) }
to a model with Hodge numbers $(h_{21},h_{11})=(69,5)$.
We choose trivial dressing $(\Delta_j=0, \omega=0,\omega_\alpha=0)$ and
introduce four D-branes of type
\eqn\fistb{ |S^a_0;\prod_j (L^a_j,M^a_j,S^a_j)\ra= |0;(1,-7,0)(0,-6,0)(3,-7,0)(0,-4,0)(0, 2, 0)\ra }
and their $\Omega$ images.
From the annulus and M\"obius strip amplitude we learn that this brane does not need
to be resolved and that it gives rise to a $U(4)$ gauge symmetry.
Next we introduce  stacks of two D-branes of type
\eqn\fistb{ |S^b_0;\prod_j (L^b_j,M^b_j,S^b_j)\ra= |0;(0,-6,0)(0,-6,0)(3,-7,0)(3,-5,0)(0,2, 0)\ra }
and 
\eqn\fistc{ |S^c_0;\prod_j (L^c_j,M^c_j,S^c_j)\ra= |0;(0,-6,0)(0,-4,0)(3,-7,0)(3,-7,0)(0, 2, 0)\ra. }
These D-branes turn out to be not single objects but are made of two fractional branes each.
Moreover, each one gives rise to a gauge symmetry $SP(2)\times SP(2)\simeq SU(2)\times SU(2)$. 
One can show that all
six tadpoles do vanish for this configuration and that the intersection numbers give rise
to the chiral spectrum as shown in Table 1
\vskip 0.8cm
\vbox{ \centerline{\vbox{ \hbox{\vbox{\offinterlineskip
\def\tablespace{height2pt&\omit&&
 \omit&\cr}
\def\tablerule{\tablespace\noalign{\hrule}\tablespace}

\hrule\halign{&\vrule#&\strut\hskip0.2cm\hfill #\hfill\hskip0.2cm\cr
& deg. &&  $U(4)\times SU(2)\times SU(2)\times SU(2)\times SU(2)$  &\cr
\tablerule
& $2$ &&  $({\bf 4}, {\bf 2}, 1 ,1,1)$  &\cr
\tablespace
& $2$ &&  $({\bf 4}, 1, {\bf 2} ,1,1)$  &\cr
\tablespace
& $2$ &&  $(\o{\bf 4}, 1,1, {\bf 2} ,1)$  &\cr
\tablespace
& $2$ &&  $(\o{\bf 4}, 1,1,1, {\bf 2})$  &\cr
\tablespace
}\hrule}}}} 
\centerline{ \hbox{{\bf
Table 1:}{\it ~~ massless chiral matter spectrum}}} } 
\vskip 0.5cm
\noindent
Therefore this Gepner model orientifold gives rise to a two  generation 
supersymmetric PS-like model. It is beyond the scope of this paper to dwell
upon the phenomenological features of this model. 
We consider this merely as a hint that supersymmetric Standard-like models are likely to
be  contained in the enormously huge  
class of Gepner model orientifolds \foot{Recently, based on 
an extensive computer search the authors of \DijkstraYM\ claimed to have found 
three generation Standard-like models.}.

\vskip 1cm
\centerline{{\bf Acknowledgements}}\pano
The work of R.B. is supported  by PPARC and T.W. is grateful to DAAD and
EPSRC.

\vfill\eject
\listrefs

\bye
\end